\documentclass[apsrev4-1,prb,superscriptaddress,longbibliography,twocolumn,10pt,aps]{revtex4-2}
\usepackage{graphicx} 
\usepackage{xcolor}
\usepackage{textpos}
\usepackage{hyperref}
\usepackage{amsmath}
\usepackage{bm}
\usepackage{physics}
\usepackage{cleveref}
\newcommand{\mrm}{\mathrm}

\newcommand{\up}{\uparrow}

\usepackage{multirow}
\begin{document}

\title{Ground-state phases of $S = 1/2$ Heisenberg models on the body-centered cubic lattice}
\author{Rajah P. Nutakki}
\affiliation{CPHT, LIX, CNRS, Inria, École Polytechnique, Institut Polytechnique de Paris, 91120 Palaiseau, France.}
\affiliation{Coll\`ege de France, Universit\'e PSL, 11 place Marcelin Berthelot, 75005 Paris, France}
\author{Filippo Vicentini}
\affiliation{CPHT, LIX, CNRS, Inria, École Polytechnique, Institut Polytechnique de Paris, 91120 Palaiseau, France.}
\affiliation{Coll\`ege de France, Universit\'e PSL, 11 place Marcelin Berthelot, 75005 Paris, France}
\begin{abstract}
Simulating low-temperature properties of three-dimensional frustrated quantum magnets is challenging due to the sign problem and the system sizes required to mitigate substantial finite-size effects.
However, there are many experimental examples of three-dimensional crystals that could host exotic low-temperature states of matter, such as quantum spin liquids.
We calculate the ground-state phase diagrams of frustrated quantum spin models on the body-centered cubic lattice using neural quantum states.
First, we study the antiferromagnetic $J_1-J_2$ model where we find a direct first-order phase transition between Néel and collinear long-range-ordered phases at $(J_2/J_1)_c = 0.705$, consistent with previous studies.
Then, in a tetragonally-distorted variant, proposed as a minimal model of NaCa$_2$Cu$_2$(VO$_4$)$_3$, we find no evidence of a quantum paramagnetic ground state, with a first-order phase transition between Néel and chain phases at $(J_{2ab}/J_1)_c = 1.0375$.
Therefore, the ground state of the tetragonally-distorted model does not reproduce the low-temperature magnetic properties of NaCa$_2$Cu$_2$(VO$_4$)$_3$, and the inclusion of other effects is necessary to rationalize experimental observations.
\end{abstract}
\maketitle
\section{Introduction}
Finding new states of matter in solid-state materials is an important challenge, both fundamentally, in terms of understanding what phases of matter are possible and from a practical point of view for developing new technologies.
In frustrated magnetism~\cite{lacroix2011}, a key goal has been to positively identify quantum spin liquids (QSLs).
In three-dimensional crystals, there are several candidate systems~\cite{okamoto2007, chillal2020, zivkovic2021, khatua2022, sana2023}, notably the rare-earth pyrochlores~\cite{rau2019,sibille2015,sibille2018,gaudet2019,gao2019,sibille2020,smith2022,smith2025,gao2025}, displaying signatures of spin-fractionalization and emergent gauge fields.
Recent experimental work~\cite{lussier2019, alexanian2025} on vanadate garnets~\cite{bayer1965,geller1967}, in particular the absence of magnetic order, dynamic Cu$^{2+}$ moments down to $50 \: \mathrm{mK}$ and broad, dispersive features in the dynamical structure factor in NaCa$_2$Cu$_2$(VO$_4$)$_3$ (NCCVO)~\cite{alexanian2025} suggests spin models on the body-centered cubic lattice with competing exchange interactions could host QSL physics.

Identifying QSLs in solid-state materials requires accurate numerical simulations to identify specific signatures of the proposed QSL that can be compared to experiments.
However, accurately simulating the low-temperature properties of three-dimensional frustrated quantum magnets is a significant challenge in itself, due to the sign problem and the large number of variables that must be simulated to mitigate finite-size effects.
Methods that have been successfully applied include diagrammatic techniques: pseudo-(Majorana)fermion functional renormalization group (PFFRG)~\cite{reuther2010a,reuther2010b,niggemann2021,niggemann2022,muller2024} and diagrammatic Monte Carlo~\cite{kulagin2013a,kulagin2013b,huang2016}, variational approaches: DMRG~\cite{hagymasi2022}, many-variable variational Monte Carlo~\cite{tahara2008,astrakhantsev2021,pohle2023}, and cluster methods~\cite{bishop1998,gelfand2000}: 
These methods face distinct challenges, such as reaching low temperatures in the diagrammatic case, entanglement truncation for DMRG or physical bias in the ansatz for many-variable variational Monte Carlo.

Recently, advances in the use of neural networks as physically unbiased variational ansatze (neural quantum states) have obtained high-accuracy ground states for two-dimensional frustrated quantum magnets~\cite{roth2023,chen2024,rende2024,viteritti2025,duric2025}.
This approach can, in principle, be extended to three-dimensional systems, however, there are few examples in the literature~\cite{astrakhantsev2021}.

\begin{figure}[tb]
	\centering
	\includegraphics[width=0.32\columnwidth]{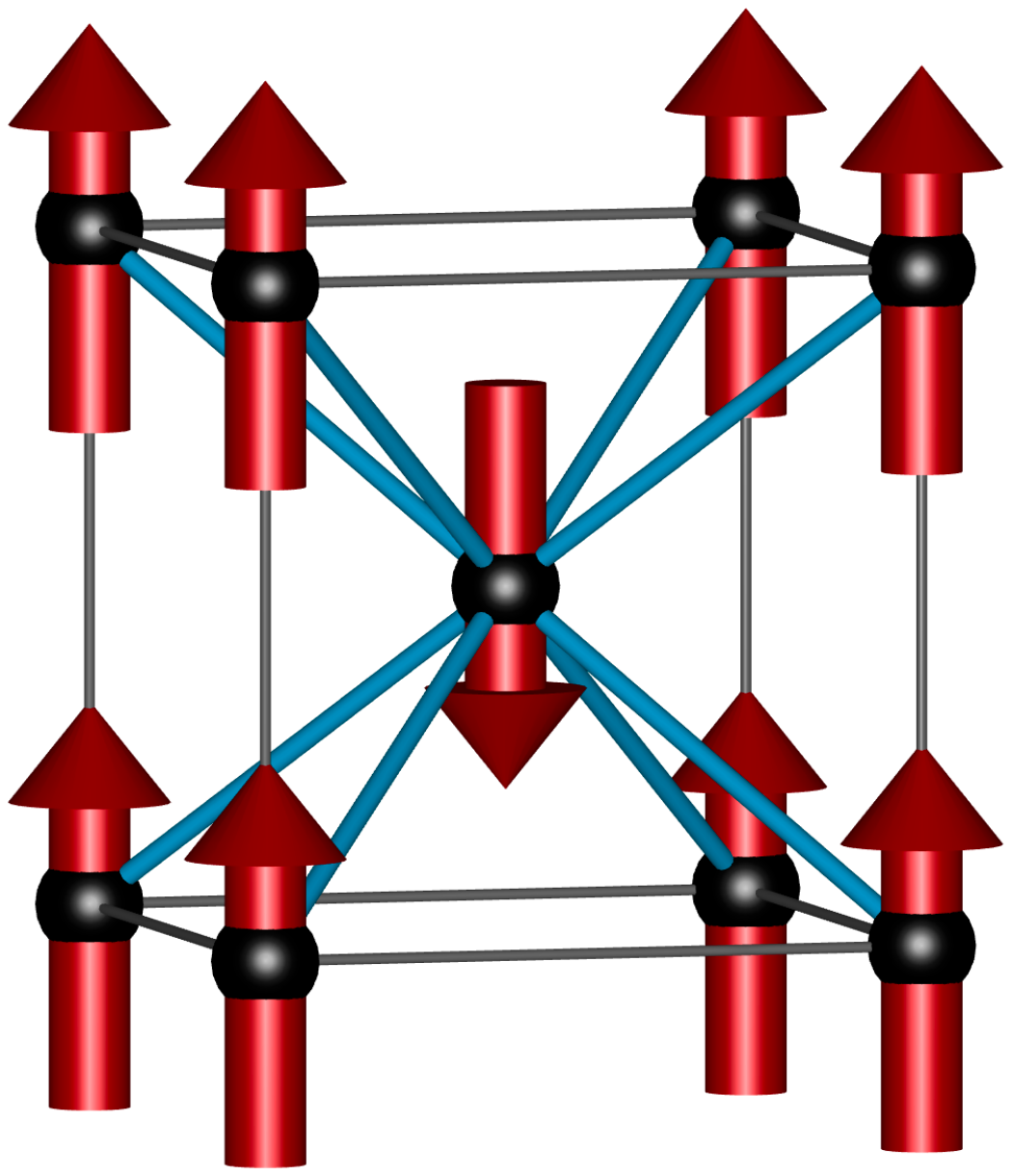}
	\includegraphics[width=0.32\columnwidth]{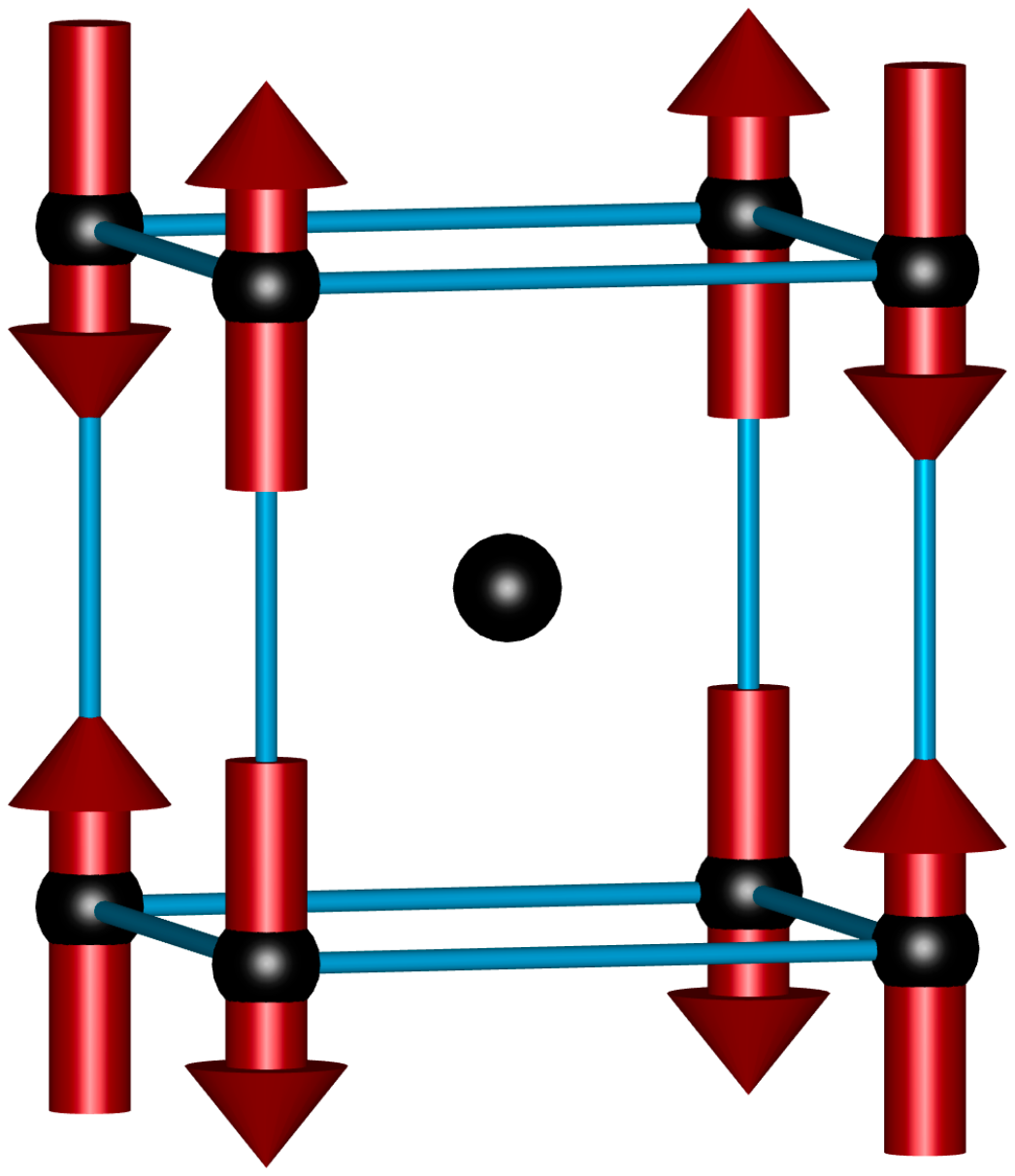}
	\includegraphics[width=0.32\columnwidth]{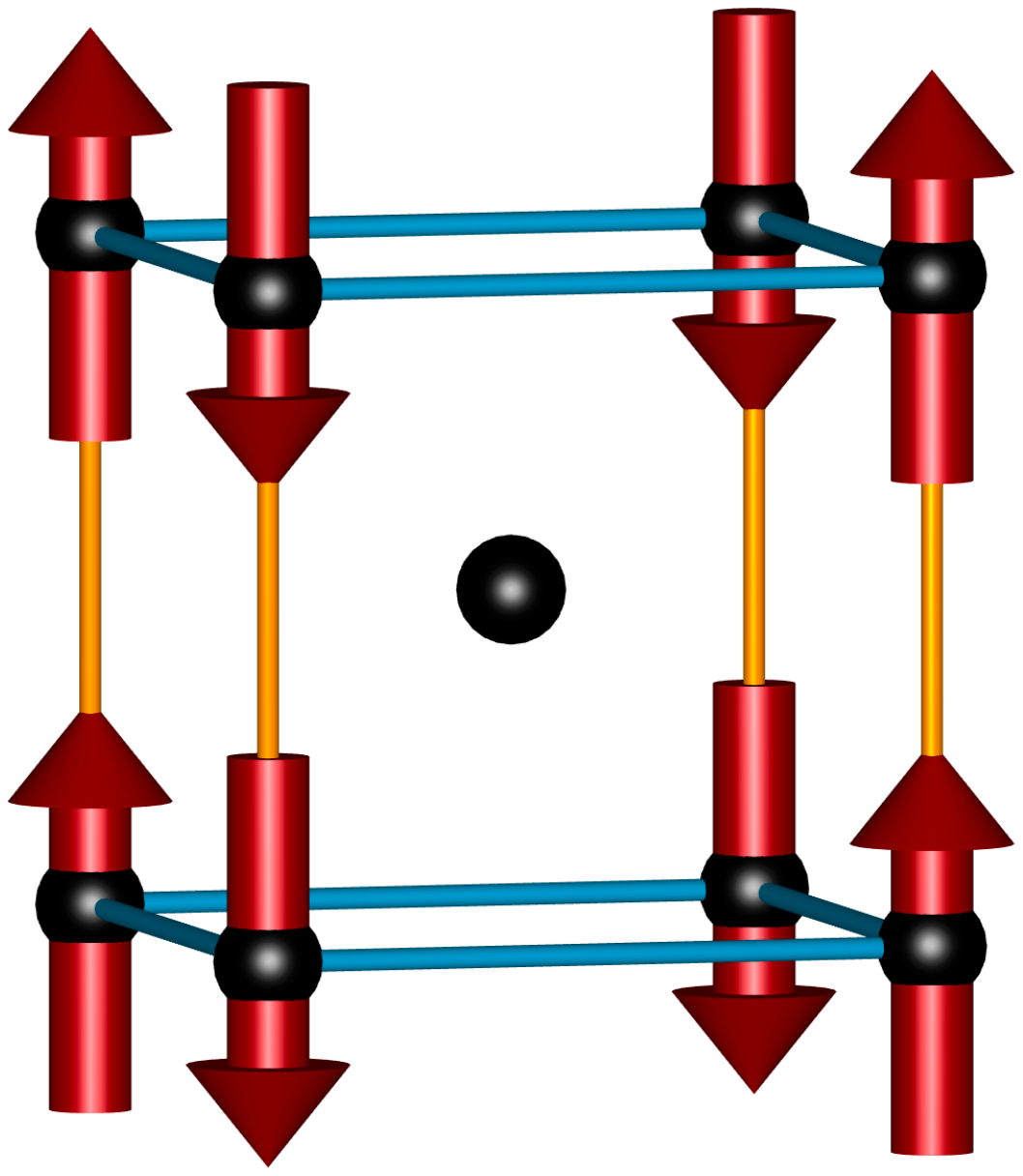}
	\begin{textblock}{1}(-0.5,-2)
		{\textbf{(a)}}
	\end{textblock}
	\begin{textblock}{1}(1.7,-2)
		{\textbf{(b)}}
	\end{textblock}
	\begin{textblock}{1}(3.8,-2)
		{\textbf{(c)}}
	\end{textblock}
	\begin{textblock}{1}(0.7,-0.3)
		{Néel}
	\end{textblock}
	\begin{textblock}{1}(2.9,-0.3)
		{Collinear}
	\end{textblock}
	\begin{textblock}{1}(5,-0.3)
		{Chain}
	\end{textblock}
	\caption{Ground states of the classical cubic and tetragonal $J_1-J_2$ models on the body-centered cubic lattice. The exchange interactions satisfied by the state are highlighted. 
		\textbf{(a)} Antiferromagnetic $J_1$ interactions are satisfied in the Néel state,  \textbf{(b)} antiferromagnetic $J_2$ interactions by the collinear state and  \textbf{(c)} antiferromagnetic $J_{2ab}$ (blue), ferromagnetic $J_{2c}$ (yellow) by the chain state.
		The sites shown make up the two-site cubic unit-cell of the body-centered cubic lattice.
	}
	\label{fig:ordered_states}
\end{figure}

In this work, we apply neural quantum states (NQS) to study the ground states of frustrated $S = 1/2$ Heisenberg models on the body-centered cubic lattice, simulating lattices of up to $288$ spins.
First, we study the (cubic-symmetric) antiferromagnetic $J_1-J_2$ model, 
\begin{equation}
		\hat{H}_{\mathrm{cub}} = J_1 \sum_{\langle ij \rangle} \hat{\mathbf{S}}_i \cdot \hat{\mathbf{S}}_j + J_{2} \sum_{\langle \langle ij \rangle \rangle} \hat{\mathbf{S}}_i \cdot \hat{\mathbf{S}}_j,
		\label{eq:ham_cubic}
\end{equation}
where $J_1$ ($J_2$) couples first (second) nearest neighbors, as illustrated in Fig.~\ref{fig:ordered_states}.
Our large-scale variational calculations find a first-order phase transition between long-range-ordered Néel and collinear phases at $(J_2/J_1)_c = 0.705 \pm 0.005$,  slightly higher than the location of the phase transition in the corresponding classical Ising model at $(J_2/J_1)_c = 0.67$, and consistent with previous results~\cite{schmidt2002,oitmaa2004,majumdar2009,pantic2014,farnell2016,ghosh2019} .

We then turn to an experimentally relevant tetragonally-symmetric variant, 
\begin{equation}
	\hat{H}_{\mathrm{tet}} = J_1 \sum_{\langle ij \rangle} \hat{\mathbf{S}}_i \cdot \hat{\mathbf{S}}_j + J_{2c} \sum_{(ij)\in 2c} \hat{\mathbf{S}}_i \cdot \hat{\mathbf{S}}_j + J_{2ab} \sum_{(ij)\in 2ab} \hat{\mathbf{S}}_i \cdot \hat{\mathbf{S}}_j, 
	\label{eq:ham_tetragonal}
\end{equation}
where the $J_2$ term is split into couplings along the $z$ ($J_{2c}$), and $x,y$-directions ($J_{2ab}$), see Fig.~\ref{fig:ordered_states}c.
This was proposed as a minimal model of the low-temperature magnetic properties of NCCVO with $J_1 = 1, J_{2c} = -4.9$ and $J_{2ab} = 1.7$~\cite{alexanian2025}.

In a classical Ising model of Eq.~\ref{eq:ham_tetragonal}, with $J_{2c}$ ferromagnetic and $J_1,J_{2ab}$ antiferromagnetic, one expects a phase transition between Néel and chain phases at $J_{2ab}/J_1 = 1$.
An intermediate quantum spin liquid phase between these phases in the quantum model would provide a qualitative explanation for the low-temperature experimental measurements on NCCVO.
We study the ground-state phase diagram of this model for $J_1 = 1$, $J_{2c} = -4.9$ and varied $J_{2ab}$ to investigate this possibility.

We do not find evidence for a quantum paramagnetic ground state, only finding long-range-ordered Néel and chain states in the parameter regime studied, with a first-order phase transition between the two at  $J_{ab}= 1.0375\pm0.0125$.
This suggests the model is missing a key feature to describe the low-temperature properties of NCCVO, possibly the effect of further neighbor couplings~\cite{ghosh2019}.
Furthermore, our work demonstrates how NQS can be used to study the low-energy properties of frustrated three-dimensional quantum Hamiltonians with hundreds of spins, assisting in the interpretation of experimental observations by providing ground-state phase diagrams of relevant minimal models.
\section{Method}
We use variational Monte Carlo with a factored vision transformer~\cite{viteritti2023} wavefunction adapted to three-dimensional systems. 
The network hyperparameters are: depth $4$, feature dimension $64$, $8$ heads and a factored-attention that spans all patches, giving a total parameter count of $\sim 10^5$ parameters.
The wavefunction is, by construction, projected to a specific momentum  $\mathbf{k}$ following the approach discussed in Ref.~\cite{nutakki2025}.
Point group symmetries are enforced by quantum number projection~\cite{tahara2008, morita2015a, nomura2021}, such that the wavefunction is symmetrized to an irreducible representation (irrep) of the lattice space group.
We restrict the calculation to the $\sum_i S^z_i = 0$ sector by a magnetization-conserving sampling scheme. 
We simulate periodic $(L_x, L_y, L_z)$ clusters of the cubic, two-site, unit cell, with lattice constant $a = 1$ (Fig.~\ref{fig:ordered_states}), and minSR~\cite{chen2024,rende2024} to optimize the variational wavefunction.

To elucidate ground-state properties of the optimized wavefunctions, we compute the static structure factor
\begin{equation}
    S(\mathbf{q}) = \frac{1}{N} \sum_{i,j} e^{i\mathbf{q}\cdot(\mathbf{r}_i - \mathbf{r}_j)} \langle \hat{\mathbf{S}}_i \cdot \hat{\mathbf S_j}\rangle
    \label{eq:structure_factor}
\end{equation}
and associated correlation ratios~\cite{kaul2015,pujari2016,nomura2021j1j2},
\begin{equation}
	R(\mathbf{Q}) = 1 - \frac{S(\mathbf{Q}+\delta \mathbf q)}{S(\mathbf Q)},
\end{equation}
where
\begin{eqnarray}
&\mathbf{Q}_{\mathrm{Neel}} = (2\pi, 2\pi, 2\pi),\nonumber \\
&\mathbf{Q}_{\mathrm{Collinear}} = (\pi, \pi, \pi), \\
\label{eq:Q}
&\mathbf{Q}_{\mathrm{Chain}} = (\pi, \pi, 0), \nonumber
\end{eqnarray}
are the ordering wavevectors of the respective ordered states (see Fig.~\ref{fig:ordered_states}), $\mathbf{r}_i$ the position of the lattice site $i$, $\delta \mathbf{q} = (2\pi/L_x, 0, 0)$ (a minimum step in momentum space), and the expectation value $\langle ... \rangle$ is estimated from Monte Carlo sampling of the optimized wavefunction.
The correlation ratio allows us to diagnose whether the ground state is an ordered or disordered phase in the thermodynamic limit.
It increases with system size in a magnetically-ordered phase and decreases in a magnetically-disordered phase. 
\begin{figure*}[t]
	\centering
	\includegraphics[width=\textwidth]{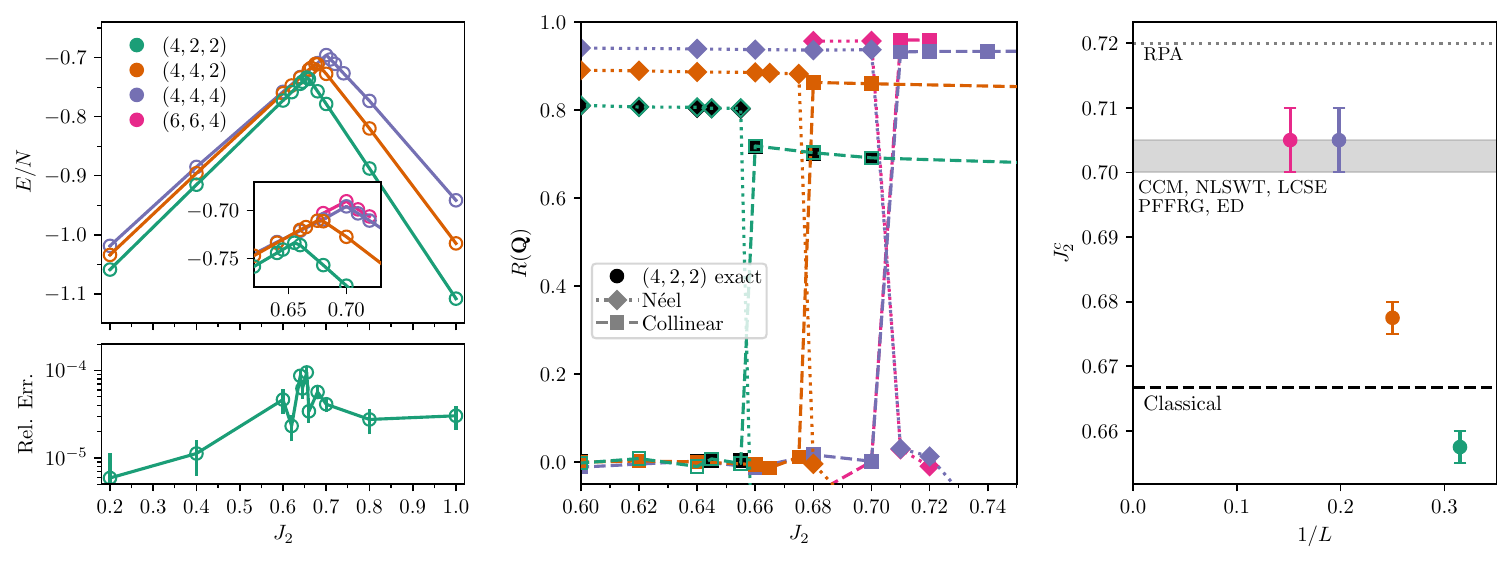}
	\begin{textblock}{1}(-0.4,-3.7)
		{\textbf{(a)}}
	\end{textblock}
	\begin{textblock}{1}(4.0,-3.7)
		{\textbf{(b)}}
	\end{textblock}
	\begin{textblock}{1}(8.7,-3.7)
		{\textbf{(c)}}
	\end{textblock}
	\caption{\textbf{(a)} (Top) Variational energy per site for the cubic-symmetric model. (Bottom) Relative error of variational energies compared to exact diagonalization for the $(4,2,2)$ lattice. Error bars (smaller than point size in the top panel) are the Monte Carlo errors when computing expectation values from the optimized wavefunction.
		\textbf{(b)} Correlation ratios for Néel and collinear orders, showing a first-order transition between the two with a critical point, $J_2^c$, that has a strong system size dependence for lattices with minimum side length $2$.
		The correlation ratios increase with system size in the respective phases, indicating a long-range-ordered phase in the thermodynamic limit.
		Exact values computed from exact diagonalization are shown.
		\textbf{(c)} The change in $J_2^c$ with system size where $L = N^{1/3}$. Error bars are given by the spacing of $J_2$ points for which we obtain converged variational solutions. 
		Our values of $J_2^c$ for system sizes with a minimum side length of $4$ are consistent with values in the literature obtained via other methods~\cite{schmidt2002,oitmaa2004,majumdar2009,farnell2016,ghosh2019}(gray band), which are well-approximated by the RPA value~\cite{pantic2014}.
	}
	\label{fig:undistorted_fig}
\end{figure*}
\section{Results and Discussion}
\subsection{Cubic Model}
First, we study the ground-state phase diagram of the cubic $J_1$-$J_2$ model.
To benchmark the method we compute the variational energies of our optimized wavefunction for a $N= 32$, $(L_x, L_y, L_z) = (4,2,2)$ lattice.
As shown in Fig.~\ref{fig:undistorted_fig}, the relative errors against the exact ground-state energy computed by exact diagonalization are $\lesssim 10^{-4}$ for all values of $J_2$.
We also perform calculations for $N = 64 (4,4,2)$, $N = 128 (4,4,4)$ and $N = 288 (6,6,4)$ lattices.
For all system sizes and phases of the model we find that the ground state is in the $\mathbf{k} = (0,0,0)$, trivial point group irrep of the space group (see Appendix~\ref{app:symmetry}).

Results for the ground-state energy and correlation ratios are presented in Fig.~\ref{fig:undistorted_fig}.
The energy is linear on both sides of the critical point, $J_2^c$, with the resulting discontinuity in $\frac{\partial E}{\partial J_2}|_{J_2^c}$ characteristic of a first-order phase transition.
Similarly, $R(\mathbf{Q})$ is discontinuous at $J_2^c$ and increases with system size within the respective phases, indicating a long-range-ordered phase in the thermodynamic limit.
Therefore, in agreement with previous studies~\cite{schmidt2002,oitmaa2004,majumdar2009,pantic2014,farnell2016,ghosh2019}, we find a direct first-order transition between the ordered ground states of the cubic $J_1-J_2$ model.

For system sizes with minimum side length, $L_{\mathrm{min}}  = 2$, we observe a significant size-dependence of $J_2^c$, in contrast to the exact diagonalization study of Ref.~\cite{schmidt2002}, where the observed system size dependence is small.
This is likely because for the lattices we use with a side length of $2$ and periodic boundaries, $J_2$ bonds couple the same sites twice along the short directions.
Nevertheless for $L_{\mathrm{min}} = 4$, we do not find any significant system size dependence of $J_2^c$, suggesting that $L_{\mathrm{min}} \geq 4$ is required on this lattice to be free of severe finite-size effects.
For these system sizes we find $J_2^c = 0.705 \pm 0.005$, consistent with results from exact diagonalization (ED)~\cite{schmidt2002}, coupled-cluster method (CCM)~\cite{farnell2016}, non-linear spin wave theory (NLSWT)~\cite{majumdar2009}, linked cluster series expansions (LCSE)~\cite{oitmaa2004} and PFFRG~\cite{ghosh2019}.

Our variational calculation is correctly able to capture the ground-state physics, even around $J_2^c$ where the sign problem is most severe (both in terms of quantum Monte Carlo and complexity of the ground-state sign structure~\cite{schmidt2002} (see Appendix~\ref{app:sign_structure}) on lattices of up to $288$ spins.
Therefore in the following section we apply our variational approach to the tetragonally-distorted $J_1-J_2$ model, to investigate the possibility of a quantum paramagnetic ground state.
\subsection{Tetragonal Model}
\begin{figure*}
	\centering
	\includegraphics[width=\textwidth]{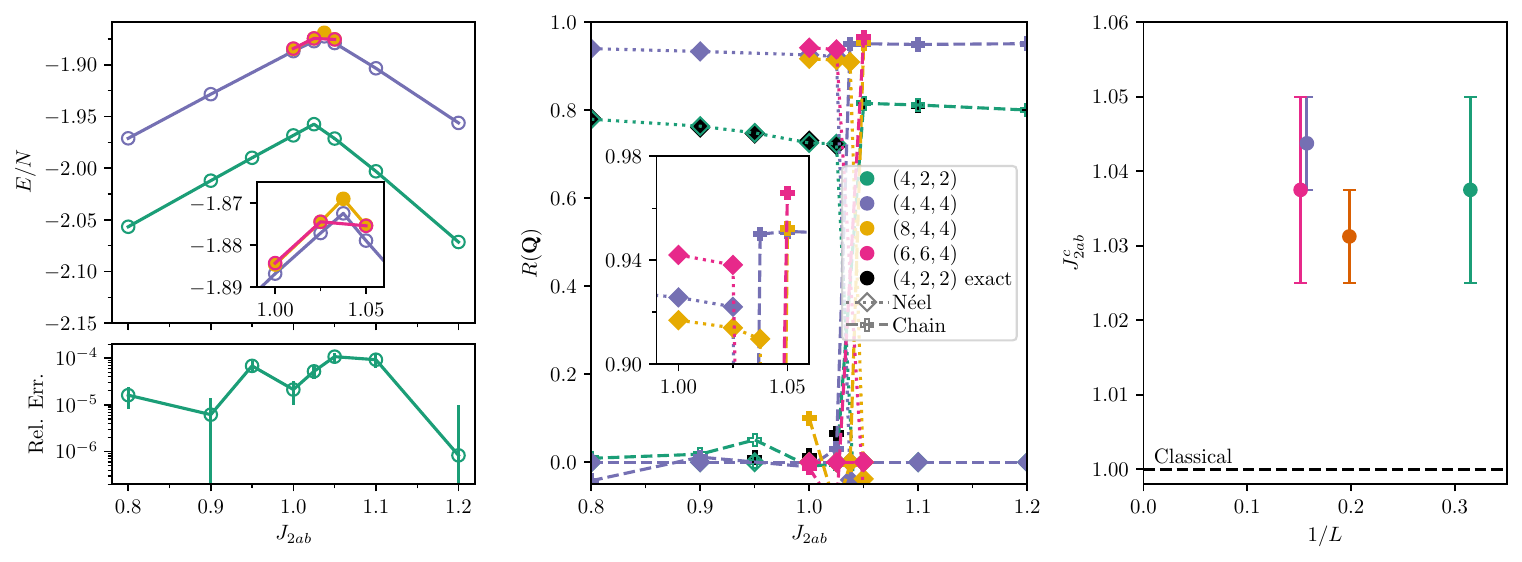}
	\begin{textblock}{1}(-0.4,-3.7)
		{\textbf{(a)}}
	\end{textblock}
	\begin{textblock}{1}(4.0,-3.7)
		{\textbf{(b)}}
	\end{textblock}
	\begin{textblock}{1}(8.7,-3.7)
		{\textbf{(c)}}
	\end{textblock}
	\caption{\textbf{(a)} (Top) Variational energy per site for the tetragonal model. (Bottom) Relative error of variational energies compared to exact diagonalization for the $(4,2,2)$ lattice. Error bars are from Monte Carlo sampling of the optimized wavefunction. \textbf{(b)} Correlation ratios for the respective ordered phases, generally showing an increase with system size, indicative of long-range-ordered phases in the thermodynamic limit, and no significant shift in the critical point, $J_{2ab}^c$. Exact values from exact diagonalization are shown.
    \textbf{(c)} The change in $J_{2ab}^c$ with system size where $L = N^{1/3}$. Error bars are given by the spacing of $J_{2ab}$ points for which we obtain converged variational solutions.
    The values are consistent with $J_{2ab}^c = 1.0375 \pm 0.0125$, higher than $J_{2ab}^c = 1$ expected for the classical Ising model.
	}
	\label{fig:tetragonal_results}
\end{figure*}
Variational results for the tetragonal $J_1-J_2$ model with $J_1 = 1, J_{2c} = -4.9$ and varied $J_{2ab}$ are shown in Fig.~\ref{fig:tetragonal_results}.
As for the cubic model, relative errors for the $(4,2,2)$ lattice are $< 10^{-4}$ for all $J_{2ab}$, and we find that the ground state is in the $\mathbf{k} = (0,0,0)$, trivial point group irrep (see Appendix~\ref{app:symmetry}).
Given the finite-size effects observed in the cubic model for $L_{\mathrm {min}} \leq 2$ we also perform calculations for $N  = 128 (4,4,4)$, $N = 256 (8,4,4)$ and $N = 288 (6,6,4)$ lattices.
We find a direct first-order phase transition between the Néel and chain long-range-ordered phases, with a discontinuity in $R(\mathbf{Q})$ and $\frac{\partial E}{\partial J_{2ab}}|_{J_{2ab}^c}$ at the transition point, $J_{2ab}^c$.
We find a small system size dependence of $J_{2ab}^c$, increasing by $1.25 \times 10^{-2}$ from the $(4,4,4)$ to $(8,4,4)$ system, both consistent with the $(6,6,4)$ value of $J_{ab}^c = 1.0375 \pm 0.0125$, with error given by the resolution of our converged calculations. 
While the correlation ratio in the Néel phase decreases slightly going from the $(4,4,4)$ to $(8,4,4)$ lattice, this is likely due to the fact that we are halving $\delta \mathbf{q}$ whilst only increasing $N$ by a factor of $2$.
Otherwise correlation ratios increase with system size in the respective ordered phases, indicative that these ordered phases survive in the thermodynamic limit.
Therefore, we find that the tetragonal $J_1-J_2$ model does not host a quantum paramagnetic ground state at the intersection of Néel and chain phases.

Previous work~\cite{ghosh2019} found that a cubic-symmetric $J_1-J_2-J_3$ model will host a quantum paramagnetic phase~\cite{ghosh2019}, nearby in parameter space to the intersection of Néel, collinear \textit{and} chain phases, where this phase competition can provide a larger low-energy manifold of states from which a quantum paramagnet can be stabilized by quantum fluctuations.
The competition in the tetragonal $J_1-J_2$ model (as in the cubic model) is only between two phases, which does not appear to be sufficient to stabilize a quantum paramagnet.

Adding $J_3$ terms with tetragonal symmetry to Eq.~\ref{eq:ham_tetragonal} with classical Ising spins, $s_i = \pm 1$, results in the following Hamiltonian
\begin{eqnarray}
	H = J_1 \sum_{\langle ij \rangle} s_i s_j + J_{2c} \sum_{(ij) \in 2c} s_is_j + J_{2ab} \sum_{(ij) \in 2ab} s_i s_j + \nonumber \\J_{3c}  \sum_{(ij) \in 3c} s_is_j + J_{3ab} \sum_{(ij) \in 3ab} s_i s_j, \hspace{2cm}
	\label{eq:ham_j3}
\end{eqnarray}
where the $J_{3c}$ term couples third neighbors with a displacement in the $z$-direction, $J_{3ab}$ third neighbors with a displacement in the $x-y$ plane. 
An intersection of the three ordered phases exists at, for example, $J_1 = 1$, $J_{2c} = -5$, $J_{2ab} = 3.5$, $J_{3c} = -1.25$, any $J_{3ab}$.
Therefore a ferromagnetic $J_{3c}$ is one mechanism by which phase competition between the three phases could be introduced, a potential pathway to a quantum paramagnetic ground state.

Our results suggest that the tetragonal $J_1-J_2$ at zero temperature is not the correct minimal model to describe the phenomena observed in experiments on NCCVO.
Further neighbor interactions, finite temperature and chemical disorder of non-magnetic Na and Ca ions are all effects that may need to be included in order to capture the experimentally observed behavior.
\section{Conclusion and Outlook}
We have applied NQS to study  the ground-state phase diagrams of both cubic and tetragonally-symmetric $J_1-J_2$ models on the body-centered cubic lattice.
In the cubic $J_1-J_2$ model we find a first-order phase transition between Néel and collinear phases at $J_2^c = 0.705\pm0.005$, in agreement with previous studies.
For the tetragonal model, recently proposed as a minimal model for the magnetic properties of NCCVO, where there is evidence for a dynamic, paramagnetic low-temperature state, we also find a first-order phase transition between Néel and chain phases at $J_{ab} = 1.0375$, with no evidence for a quantum paramagnetic phase.
This suggests that other effects will need to be included to correctly model the experimentally-observed phenomena.

Since a quantum paramagnetic phase was previously found near the intersection of Néel, collinear and chain phases~\cite{ghosh2019} we suggest that the inclusion of a ferromagnetic $J_{3c}$ could induce a similar phase competition in the tetragonal $J_1-J_2-J_3$ model, so could host a quantum paramagnetic ground state.
Future work could map out both classical and quantum ground-state phase diagrams of this model to investigate this possibility.

We have demonstrated how NQS can be used to study experimentally-relevant questions in three-dimensional frustrated magnets.
Expanding this to geometrically (rather than exchange)-frustrated three-dimensional lattices, such as the pyrochlore, could require learning more complex ground-state sign structures~\cite{westerhout2020, schurov2025}, which may not be easily done with current techniques.
Understanding how to address this issue would be an important step to broaden the applicability of NQS to frustrated magnets.
\section*{Data and code availability}
The data supporting the findings of this article are available~\cite{data}, as well as the code for our NQS calculations~\cite{code}.
\section*{Acknowledgments}
We thank F. Bert and K. Salou-Smith for useful discussions, and E. Kermarrec, F. Alet and S. Capponi for helpful discussions and comments on the manuscript.
The authors acknowledge support by the French Agence Nationale de la Recherche through the NDQM project, grant ANR-23-CE30-0018.
This work was performed using HPC
resources from GENCI-IDRIS (Grants 2025-AD010517107 and 2024-A0170515698) on the V100 and A100 partitions of Jean-Zay.
\appendix
\section{Symmetry sectors}
\label{app:symmetry}
We use a wavefunction projected to a momentum sector, $\mathbf{k}$, of the patched lattice.
Since we work with a simple-cubic Bravais lattice with patches made up of the two-site basis, the translation group is that of the simple-cubic lattice.
The wavefunction can further be projected to space-group sectors through quantum number projection~\cite{tahara2008,morita2015a,nomura2021,reh2023} over the corresponding little group characters, $\chi_g$, as well as to a spin-parity sector, via
\begin{equation}
	\Psi_{\mathbf k, L_{\mathbf{k}},\theta}(\bm \sigma) = \frac{1}{\abs{L_{\mathbf k}}} \sum_{g \in L} \chi_g \Psi_{\mathbf k, \theta}(g^{-1} \bm \sigma)
\end{equation}
and
\begin{equation}
	\Psi_{\mathbf k, L_{\mathbf{k}},P,\theta}(\bm \sigma) = \frac{1}{2}  (\Psi_{\mathbf k, L_{\mathbf k}, \theta}(\bm \sigma) \pm_P \Psi_{\mathbf k, L_{\mathbf k}, \theta}(- \bm \sigma)).
\end{equation}
In practice, we gradually introduce symmetries during the course of the optimization~\cite{viteritti2023}.

\begin{figure*}
	\includegraphics[width=14cm]{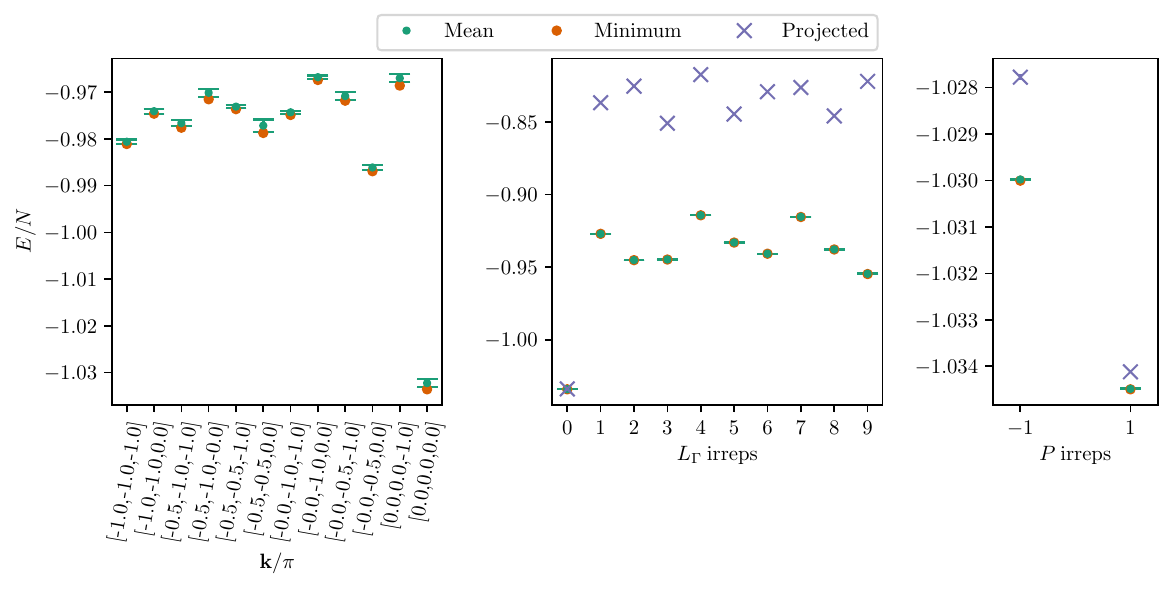}
	\begin{textblock}{1}(1,-3.6)
		{\textbf{(a)}}
	\end{textblock}
	\begin{textblock}{1}(5.2,-3.6)
		{\textbf{(b)}}
	\end{textblock}
	\begin{textblock}{1}(9,-3.7)
		{\textbf{(c)}}
	\end{textblock}
	\caption{Finding the optimal symmetry sector for the cubic model. \textbf{(a-c)} Variational energies obtained in different symmetry sectors for $(4,4,2), J_2 = 0.2$. The mean is the final energy of 5 different runs with error bars given by the standard deviation. We also plot the energy obtained by projecting the wavefunction from the previous symmetry stage without any optimization. First, \textbf{(a)}, optimizations are carried out for all momenta in the irreducible Brillouin zone. Then, \textbf{(b)}, starting from the optimized states, further optimizations/projections are carried out for all irreps of the little group of $\mathbf{k} = \Gamma = (0,0,0)$, giving us the optimal space-group sector. Finally, \textbf{(c)}, further optimizations/projections are carried out in the spin-parity antisymmetric/symmetric sectors. Since projections find the same optimal symmetry sector as running further optimizations, for $N \geq 128$ and all system sizes on the tetragonal model we only run optimizations for different momenta then project the optimized states to find the optimal symmetry sectors. }
	\label{fig:symmetry_sectors}
\end{figure*}

In order to ascertain the symmetry sector of the ground state, we perform optimizations and projections in the various sectors for each system size up to and including $(4,4,4)$ at $J_2 = 0.2, 1$ of the cubic model and $J_{2ab} = 0.2, 1.5$ of the tetragonal model.
First, we perform optimizations for all $\mathbf k$ in the irreducible Brillouin zone.
Then we project the wavefunction of the optimized solution to all corresponding little group sectors, giving us the space-group sector of the ground state.
Finally, we project the optimal space-group symmetrized wavefunction to the two spin-parity sectors.
As shown in Fig.~\ref{fig:symmetry_sectors}, we find that this projection procedure gives the same optimal symmetry sector as performing further optimization of the wavefunction after projection.
We find the optimal symmetry sector is always the $\mathbf{k} = (0,0,0)$, invariant little group irrep and spin-parity symmetric sector, although there is a quasi-degeneracy in the little group with other sectors for $J_2 = 1$ of the cubic model and $J_{ab} = 1.5$ of the tetragonal. 
\section{Patching}
\begin{figure}
	\centering
    \includegraphics[width=8cm]{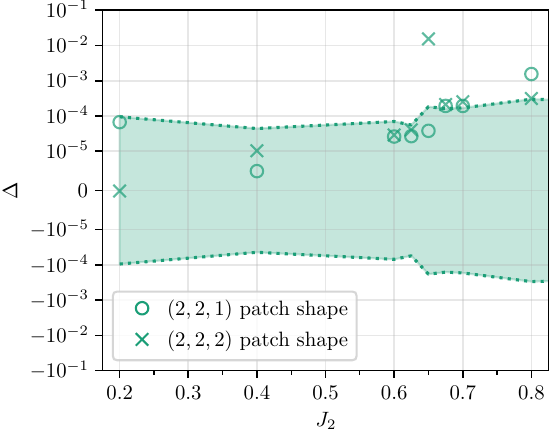}
    \caption{Comparison of the relative difference of variational energies obtained using different patchings for the (4,2,2) lattice, computed relative to a (1,1,1) patch. The shaded regions correspond to the error, due to fluctuations during the final iterations of the optimization. There is no systematic difference between the patchings.}
    \label{fig:patching}
\end{figure}

In order to compare the effects of different patchings in our variational wavefunction, we perform optimizations for the $(4,2,2)$ lattice using various patch shapes.
We define patch shapes in units $(n_x,n_y,n_z)$ of the cubic lattice vectors, comparing a $(1,1,1)$ patch (2 sites), $(2,2,1)$ patch (8 sites) and $(2,2,2)$ patch (16 sites).
To ensure the results are comparable, we optimize first without enforcing any additional translational symmetries, before enforcing all intrapatch translations, resulting in a wavefunction that is invariant under the full translation group of the lattice.
We compare the relative difference of the minimum energies obtained for each patch size,
\begin{equation}
    \Delta = \frac{E-E_{\mrm{ref}}}{E_{\mrm{ref}}},
\end{equation}
where $E_{\mrm{ref}}$ is the minimum energy obtained using patch shape (1,1,1).
A comparison across different values of $J_2$ is presented in~\Cref{fig:patching}, which shows that there is no significant, systematic, difference in variational energies obtained using different patchings.
\section{Sign Structure}
\label{app:sign_structure}
\begin{figure*}
\centering
    \includegraphics[width=5cm]{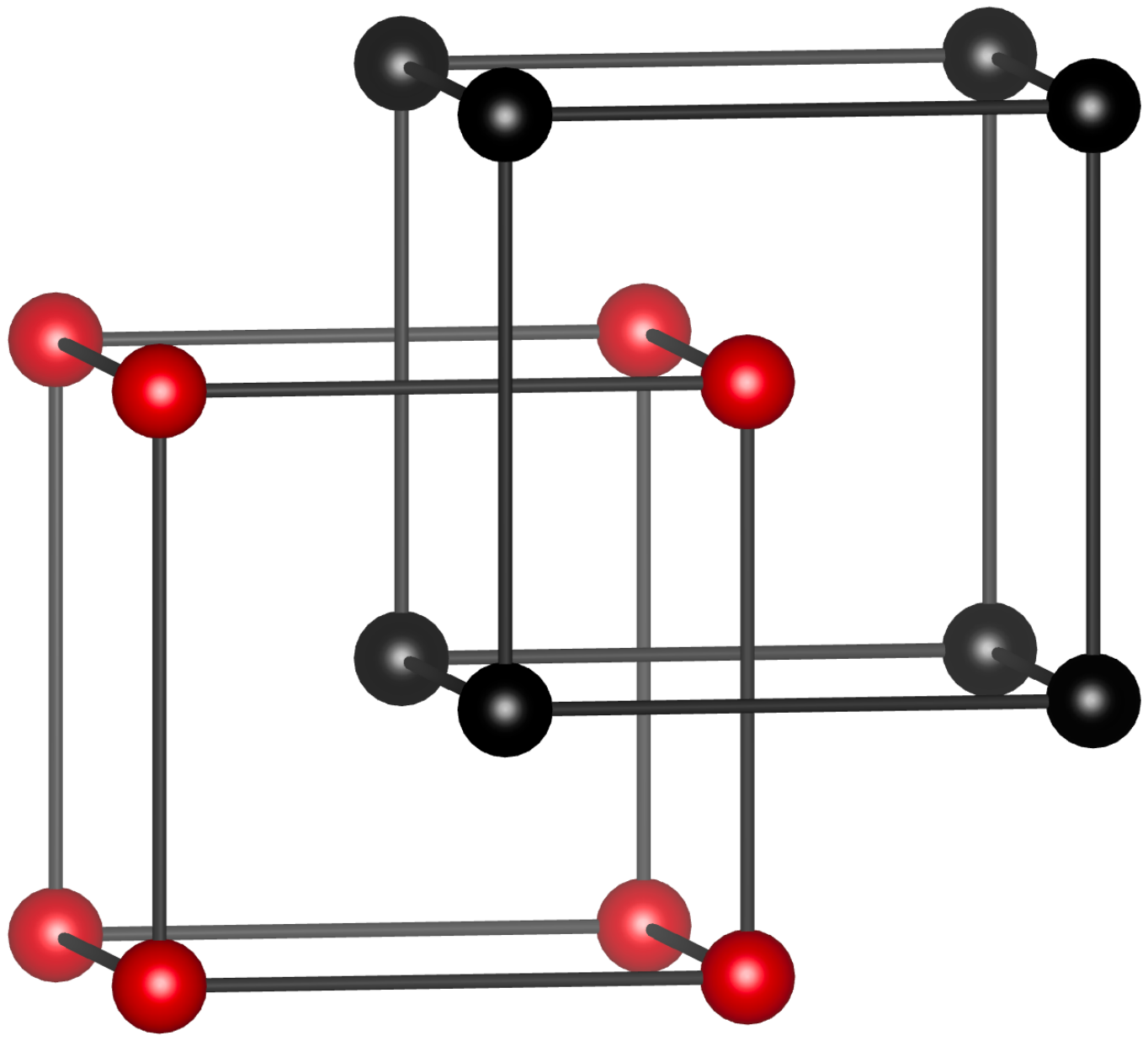}
    \hspace{1cm}
    \includegraphics[width=5cm]{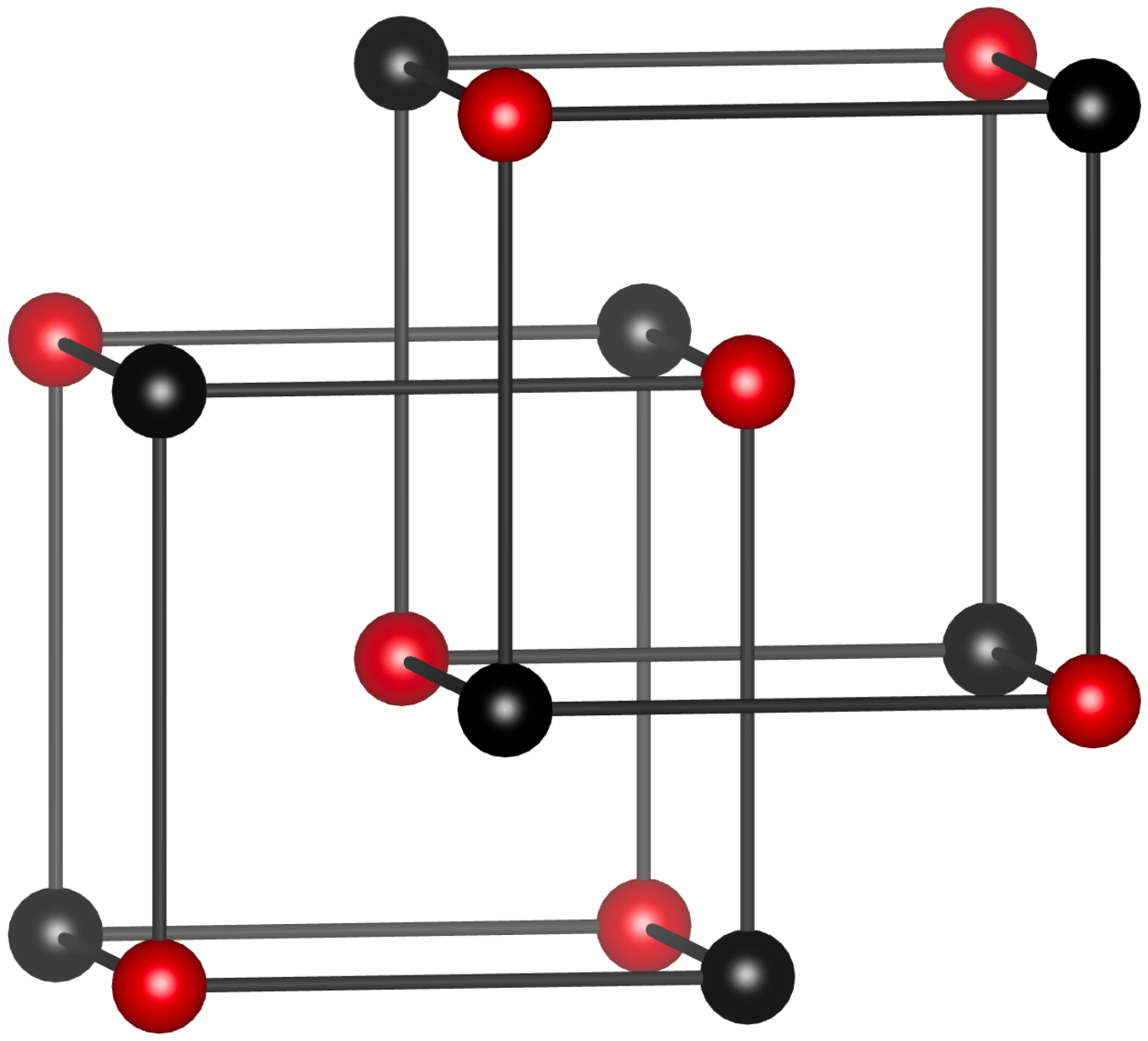}
    \hspace{1cm}
    \includegraphics[width=5cm]{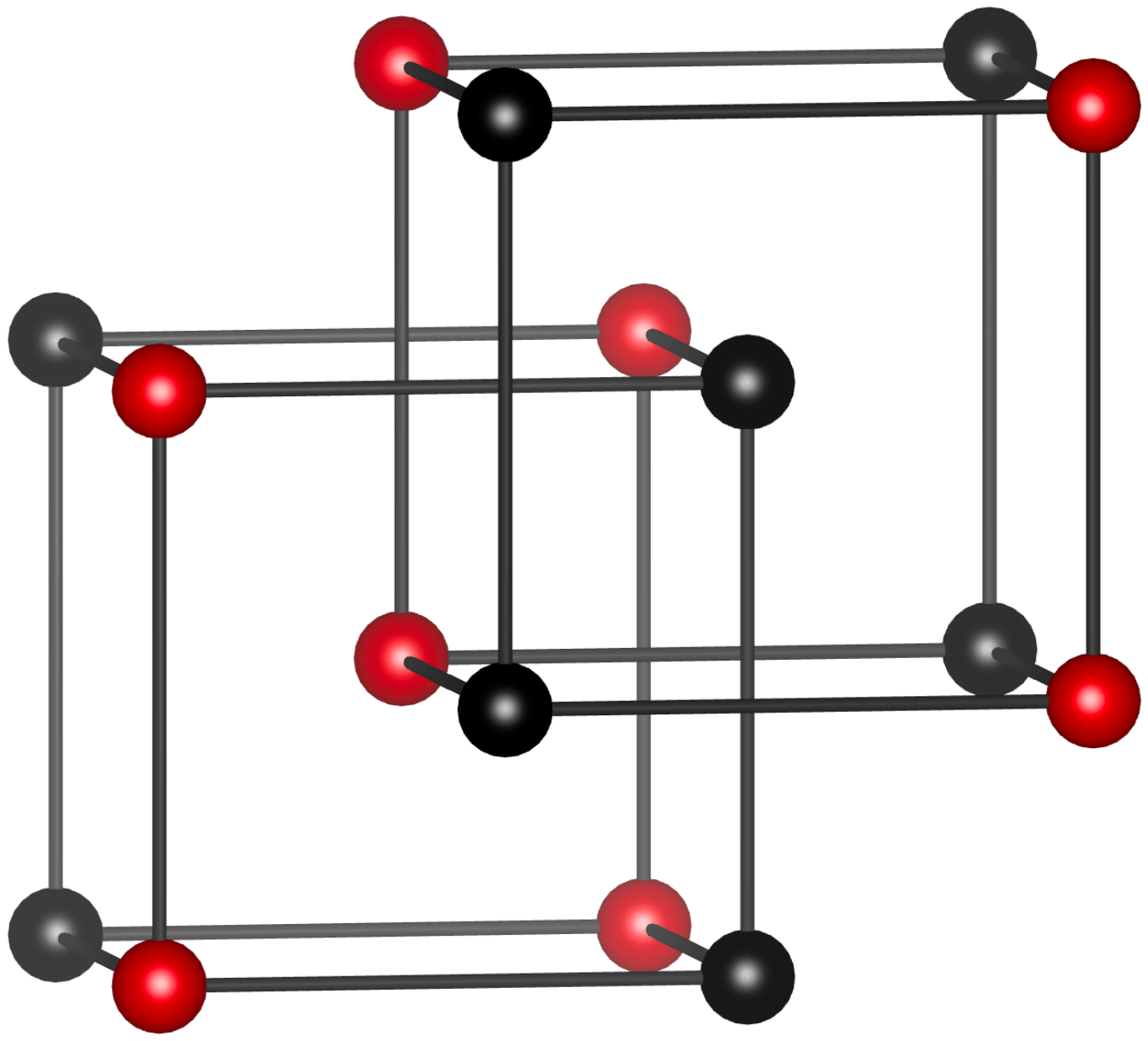}
    \label{fig:sign_sublattices}
    \begin{textblock}{1}(-0.2,-2)
		{\textbf{(a)}}
	\end{textblock}
    \begin{textblock}{1}(4.2,-2)
		{\textbf{(b)}}
	\end{textblock}
    \begin{textblock}{1}(8.8,-2)
		{\textbf{(c)}}
	\end{textblock}
    \begin{textblock}{2}(1.,0.1)
		{\Large{$J_1$}}
	\end{textblock}
    \begin{textblock}{2}(5.6,0.1)
		{\Large{$J_2$}}
	\end{textblock}
    \begin{textblock}{3}(10,0.1)
		{\Large{$J_2$ distorted}}
	\end{textblock}
    \vspace{0.5cm}
    \caption{The sublattices, $\mu$, (red) used to compute the Marshall sign rules defined in Eq.~\ref{eq:marshall_sign}. This gives the exact ground-state sign structure for Eq.~\ref{eq:ham_cubic} with (\textbf{a}) $J_1 > 0, J_2 = 0$, (\textbf{b}) $J_1 = 0, J_2 > 0$, and (\textbf{c}) for Eq.~\ref{eq:ham_tetragonal} with $J_1 = 0, J_{2c} < 0, J_{2ab} >0$.}
    \label{fig:marshall_sublattices}
\end{figure*}
The Marshall sign rules~\cite{marshall1955} (see the appendix of Ref.~\cite{moss2025} for a concise explanation) that give the exact ground-state sign structure in various limits of the Hamiltonian are
\begin{equation}
    s_{\mu}(\bm \sigma) = (-1)^{n_{\up,{\mu}}}
    \label{eq:marshall_sign}
\end{equation}
with $n_{\up,x}$ the number of spins on the sublattice $\mu$.
The various sublattices are illustrated in Fig.~\ref{fig:marshall_sublattices}.

\begin{figure*}
\centering
    \includegraphics[width=6cm]{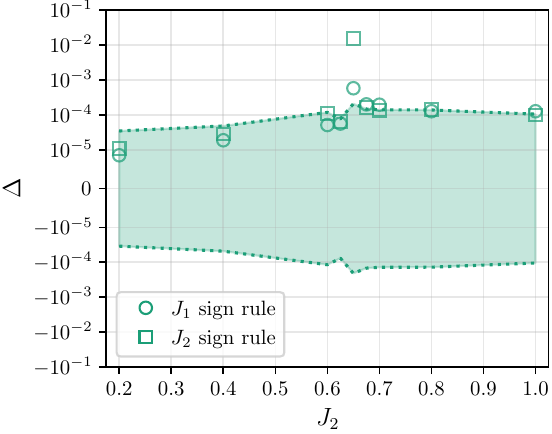}
    \hspace{1cm}
    \includegraphics[width=10cm]{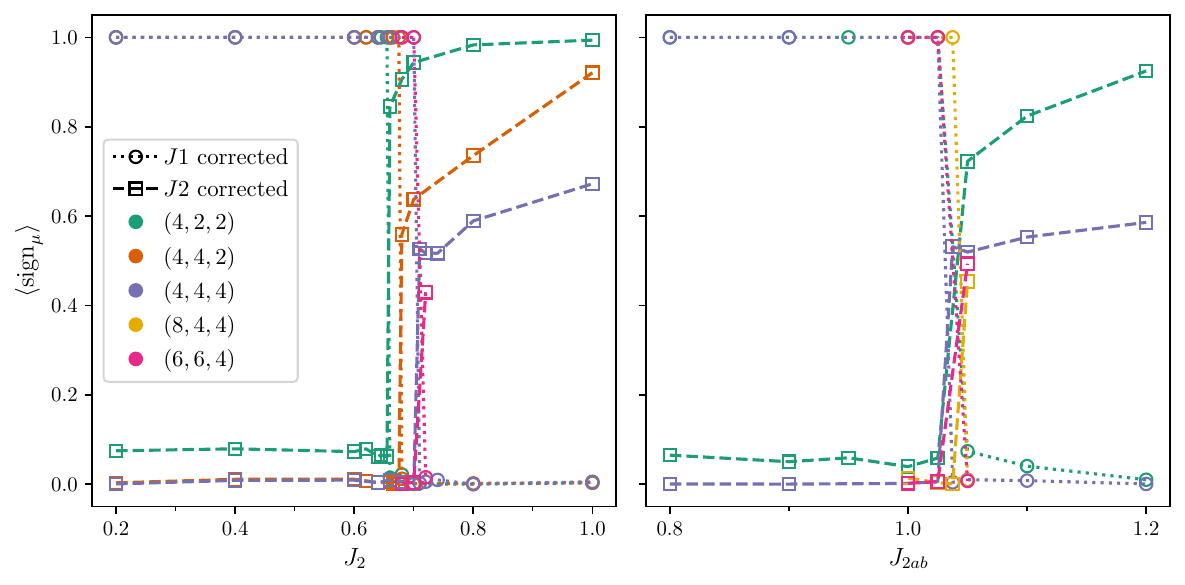}
    \begin{textblock}{1}(-0.2,-2.8)
		{\textbf{(a)}}
	\end{textblock}
    \begin{textblock}{1}(5,-2.8)
		{\textbf{(b)}}
	\end{textblock}
    \caption{\textbf{(a)} Variational energies obtained using different sign rules for the $N = 32 (4,2,2)$ lattice of the cubic-symmetric $J_1-J_2$ model for different values of $J_2$, relative to an ansatz without a sign rule. 
    The shaded regions correspond to the error, due to fluctuations during the final iterations of the optimization. The use of sign rules does not improve variational energies, even for $J_2 \gtrsim 0.65$ where relative errors to the exact solution increase.
    \textbf{(b)}
    Expectation values of the Marshall-corrected sign of variational solutions for (left) the cubic $J_1-J_2$ model and (right) tetragonal $J_1-J_2$ model.
    }
    \label{fig:signs}
\end{figure*}

For the cubic $J_1-J_2$ model, we verify whether variational energies can be improved by explicitly accounting for the Marshall sign rules in the $J_1 = 0$ and $J_2 = 0$ limits by using the ansatz
\begin{equation}
    \Psi_{\mu,\theta} (\bm \sigma) = s_{\mu}(\bm \sigma) \Psi_{\theta} (\bm \sigma),
\end{equation}
As shown in~\Cref{fig:signs}a the explicit incorporation of a sign rule does not improve variational energies for the $(4,2,2)$ lattice, suggesting the network is able to easily learn these sign rules during the optimization.

To quantify the ground-state sign structures, we compute the expectation value of the Marshall-corrected sign of the optimized wavefunction 
\begin{equation}
    \langle \mathrm{sign}_{\mu} \rangle = \sum_{\bm{\sigma}} \abs{\Psi_{\theta}(\bm{\sigma})}^2  s_{\mu}(\bm \sigma) \mathrm{sign}(\Psi_{\theta}(\bm \sigma)). 
\end{equation}
A wavefunction whose ground-state sign structure is exactly given by a Marshall sign rule yields $\langle \mathrm{sign}_{\mu}\rangle = 1$, whereas $\langle \mathrm{sign}_{\mu}\rangle = 0$ means there is no correlation between the Marshall and actual sign of the wavefunction.

Results for $\langle \mathrm{sign}_{\mu} \rangle$ for variational solutions of both cubic and tetragonal models are shown in Fig.~\ref{fig:signs}b. 
In the cubic case, as observed in exact diagonalization~\cite{schmidt2002}, across the entire Néel phase the Marshall-corrected sign is almost exactly $1$, whereas in the collinear phase the sign increases with $J_2$.
As system size is increased the average sign at fixed $J_2$ in the collinear phase decreases.
In the tetragonal case, the results are similar, but with a lower average sign in the large $J_{2ab}$ phase.
The minimum percentage of positive signs after Marshall sign correction we observe across all system sizes and parameters is $70\%$.

It is known that NQS struggles with complex sign structures~\cite{szabo2020,westerhout2020,schurov2025}, but that the ViT can learn Marshall sign rules~\cite{viteritti2025}, so the fact that Marshall sign rules give a reasonable percentage of the ground-state signs may be one factor in the success of our approach for the models we study.
\section{Structure Factors and Correlation Ratios}
\begin{figure*}
\centering
\begin{tabular}{cccc}
    &$(4,2,2)$ & $(4,4,2)$ & $(4,4,4)$ \\
  \raisebox{1.8cm}{\rotatebox{90}{$J_2 = 0.2$}}  & 
  \includegraphics[width=0.3\textwidth]{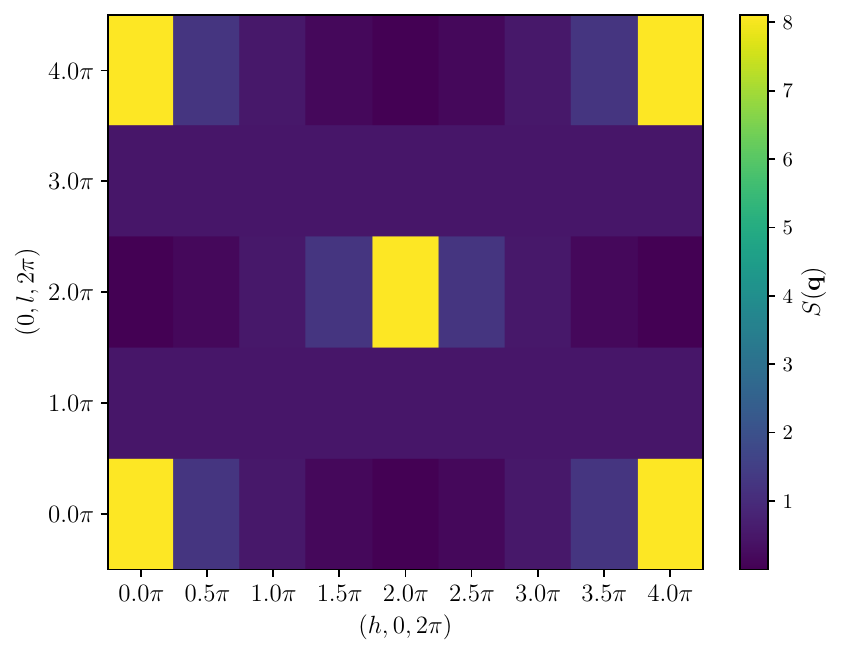} &
  \includegraphics[width=0.3\textwidth]{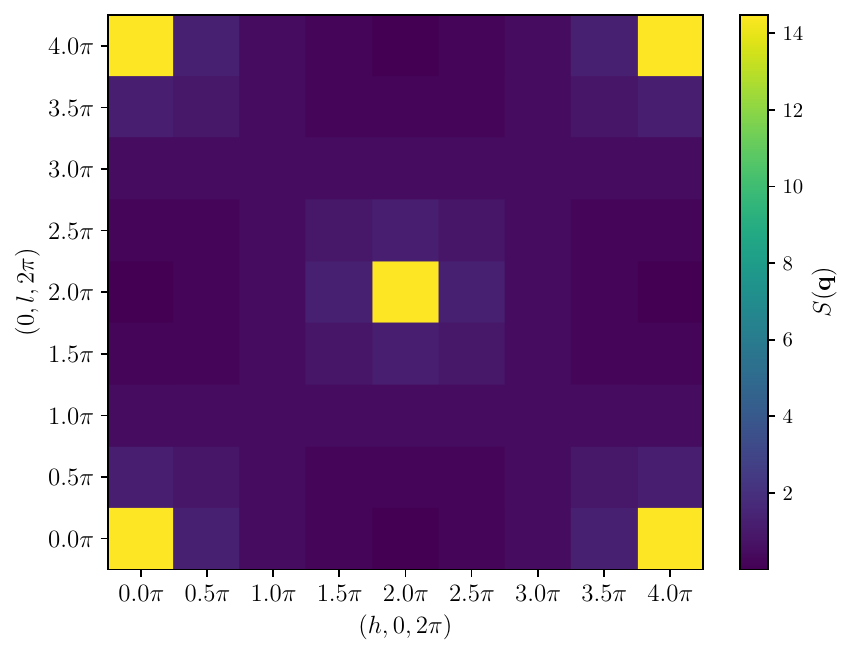} &
  \includegraphics[width=0.3\textwidth]{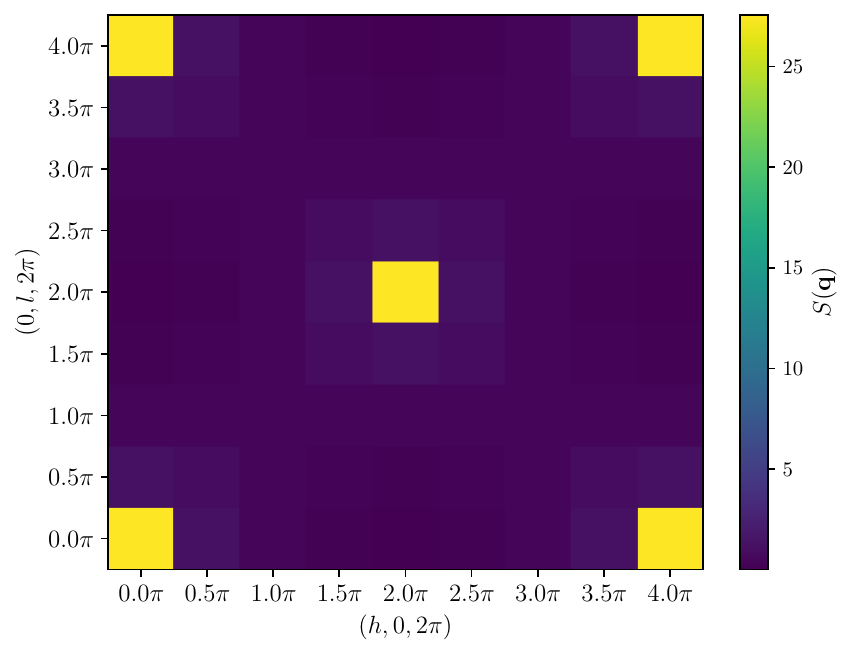} \\[6pt]
 \raisebox{1.8cm}{\rotatebox{90}{$J_2 = 0.7$}}    & 
  \includegraphics[width=0.3\textwidth]{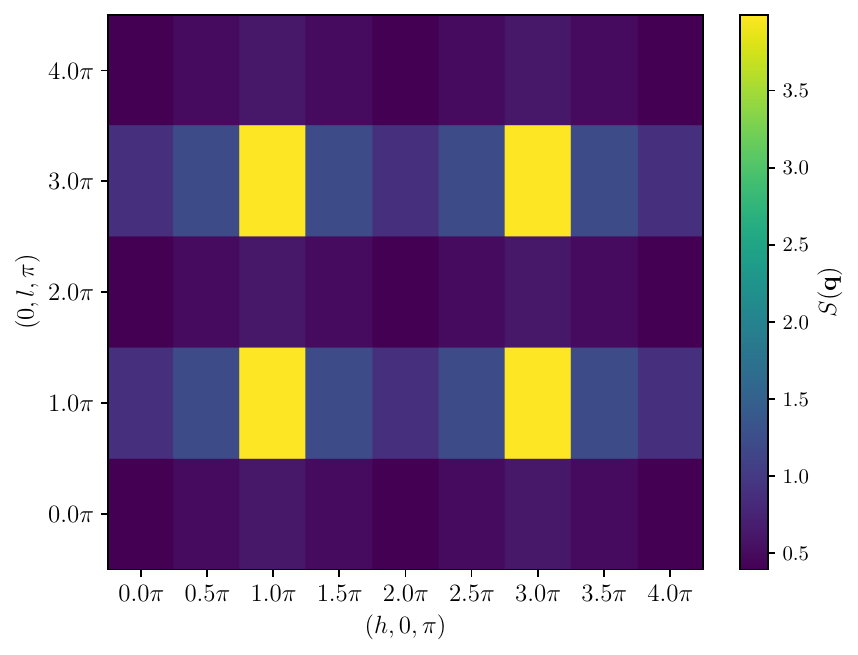} &
  \includegraphics[width=0.3\textwidth]{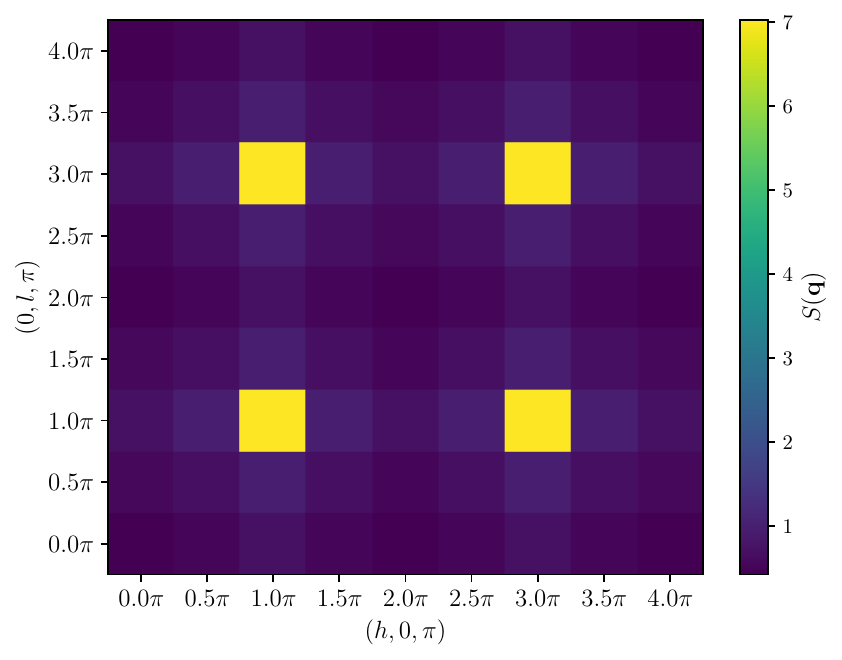} &
  \includegraphics[width=0.3\textwidth]{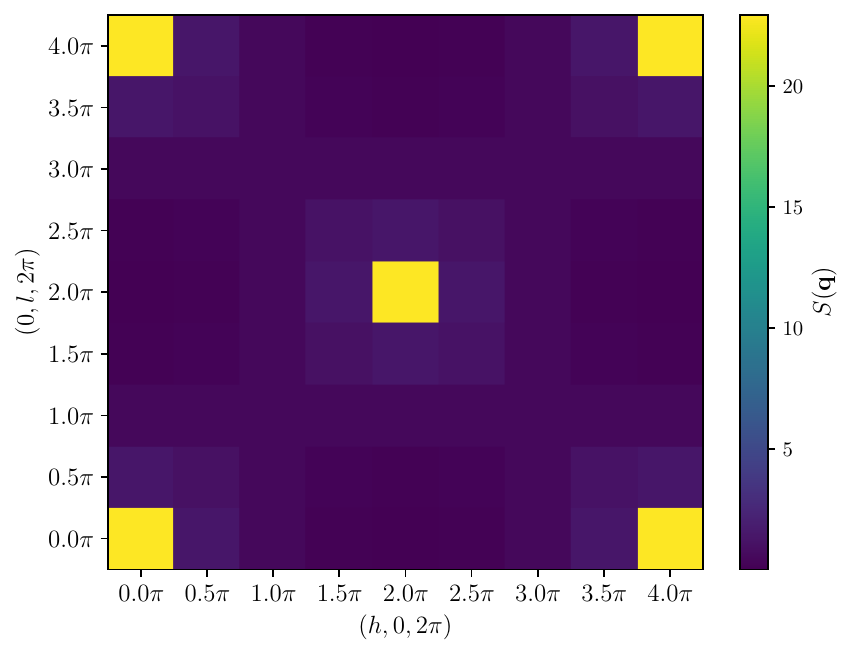} \\[6pt]
 \raisebox{1.8cm}{\rotatebox{90}{$J_2 = 1$}}  & 
  \includegraphics[width=0.3\textwidth]{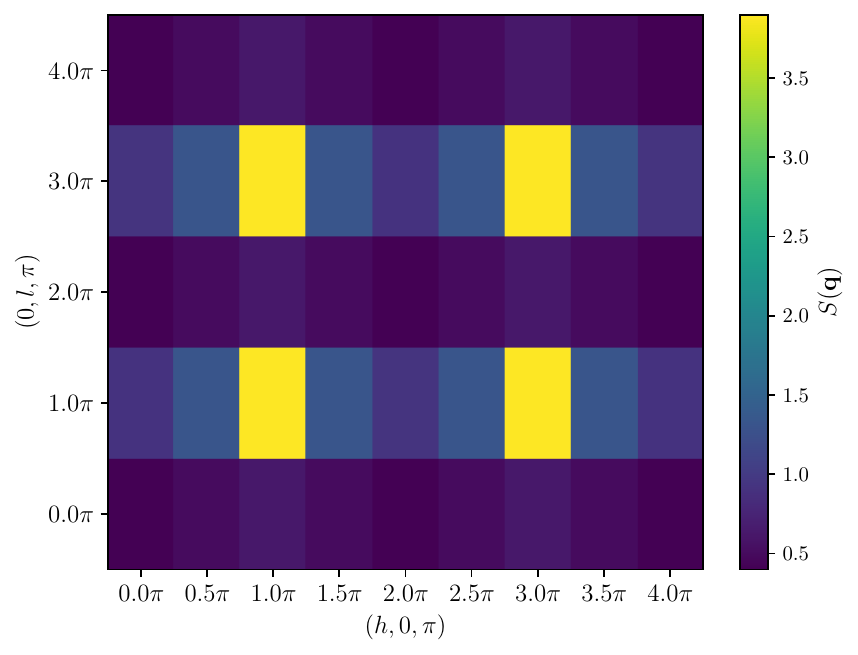} &
  \includegraphics[width=0.3\textwidth]{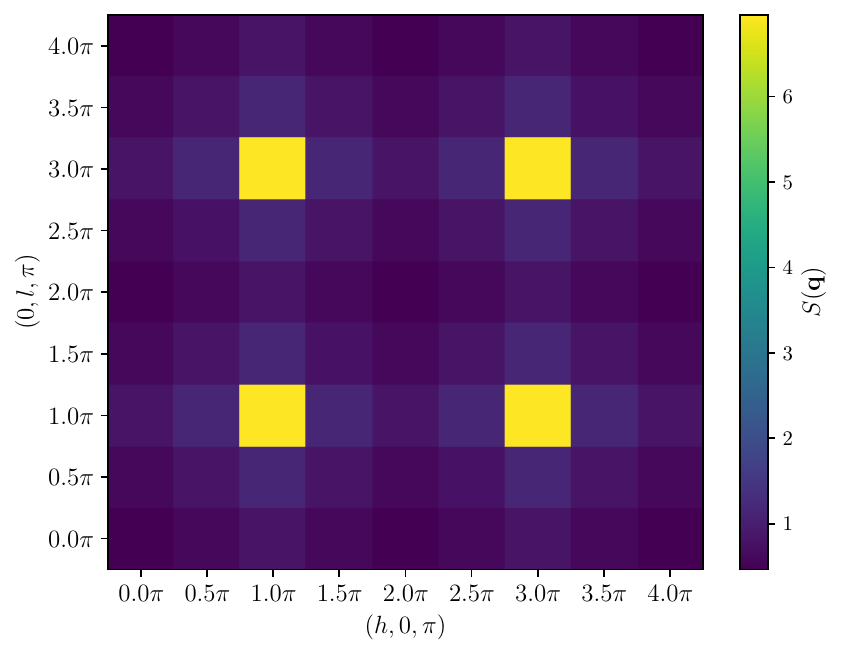} &
  \includegraphics[width=0.3\textwidth]{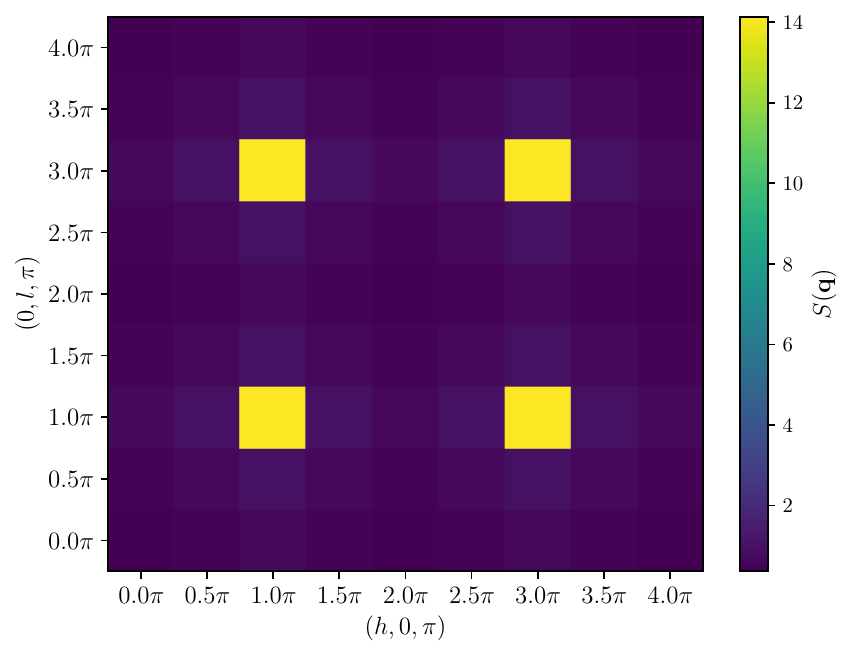}
\end{tabular}
\caption{Ground-state structure factors of the cubic $J_1-J_2$ model for various $J_2$ and system sizes, using momentum cuts that show Bragg peaks. Due to the shifting of $J_2^c$ with system size the $(4,2,2), (4,4,2), J_2 = 0.7$ ground states are in the collinear phase, while for $(4,4,4), J_2 = 0.7$ it is in the Néel phase.}
\label{fig:cubic_structure_factor}
\end{figure*}
\begin{figure*}
\centering
\begin{tabular}{cccc}
    &$(4,2,2)$ & $(4,4,4)$ & $(6,6,4)$ \\
  \raisebox{1.8cm}{\rotatebox{90}{$J_{2ab} = 1$}}  & 
  \includegraphics[width=0.3\textwidth]{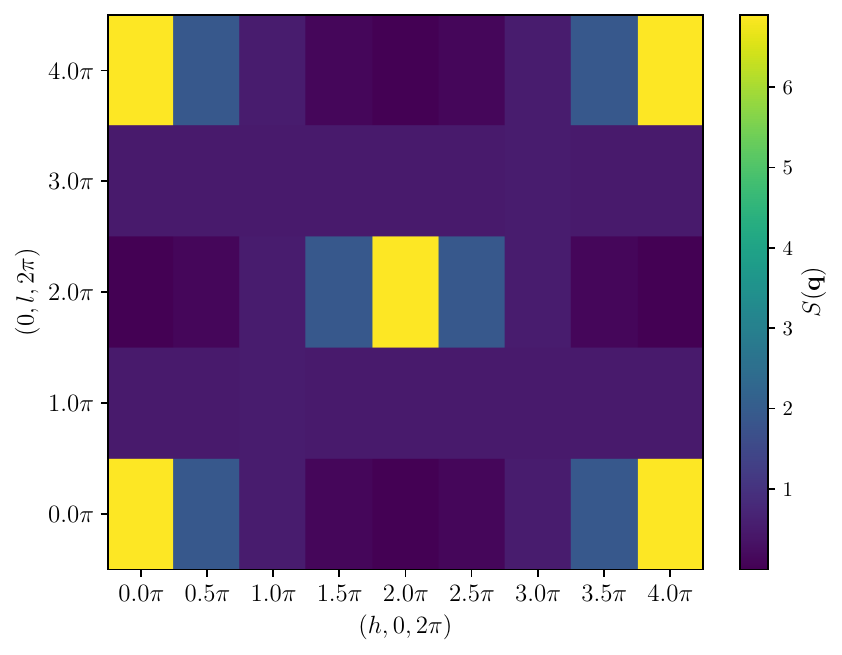} &
  \includegraphics[width=0.3\textwidth]{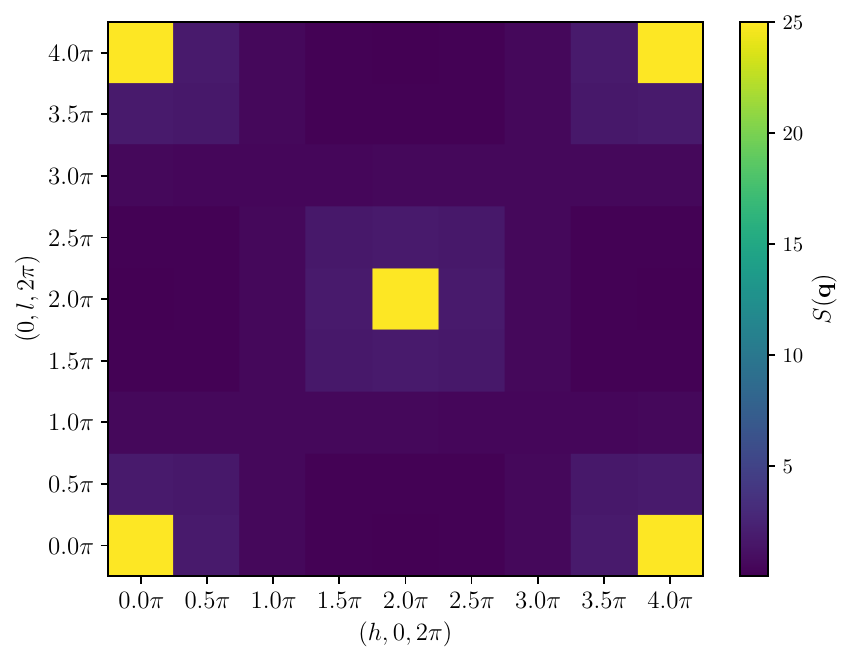} &
  \includegraphics[width=0.3\textwidth]{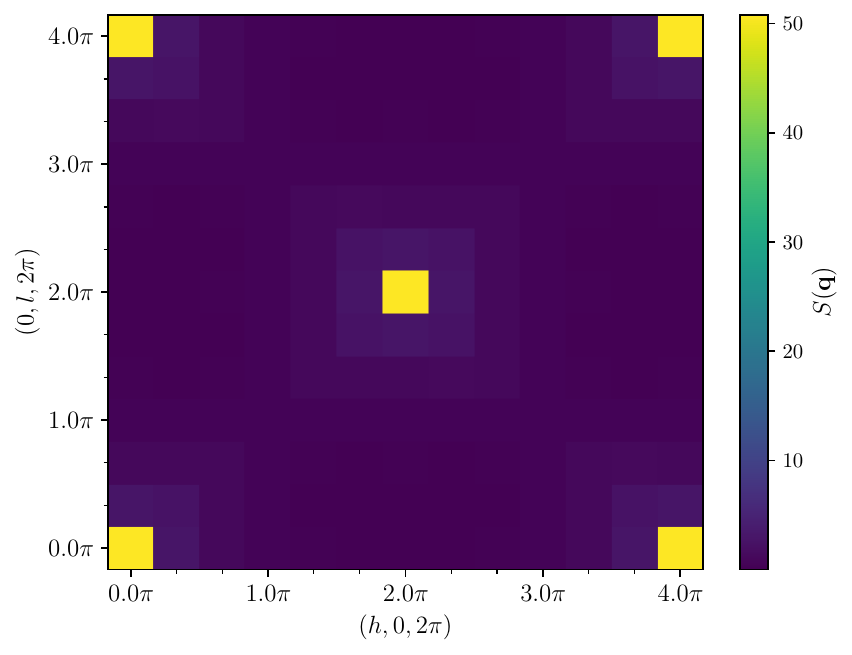} \\[6pt]
 \raisebox{1.65cm}{\rotatebox{90}{$J_{2ab} = 1.05$}} & 
  \includegraphics[width=0.3\textwidth]{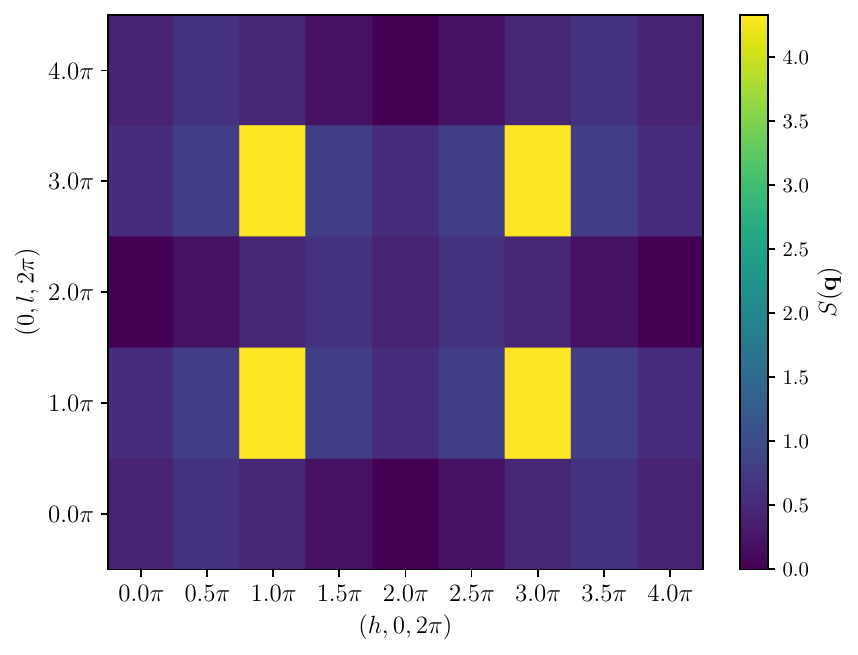} &
  \includegraphics[width=0.3\textwidth]{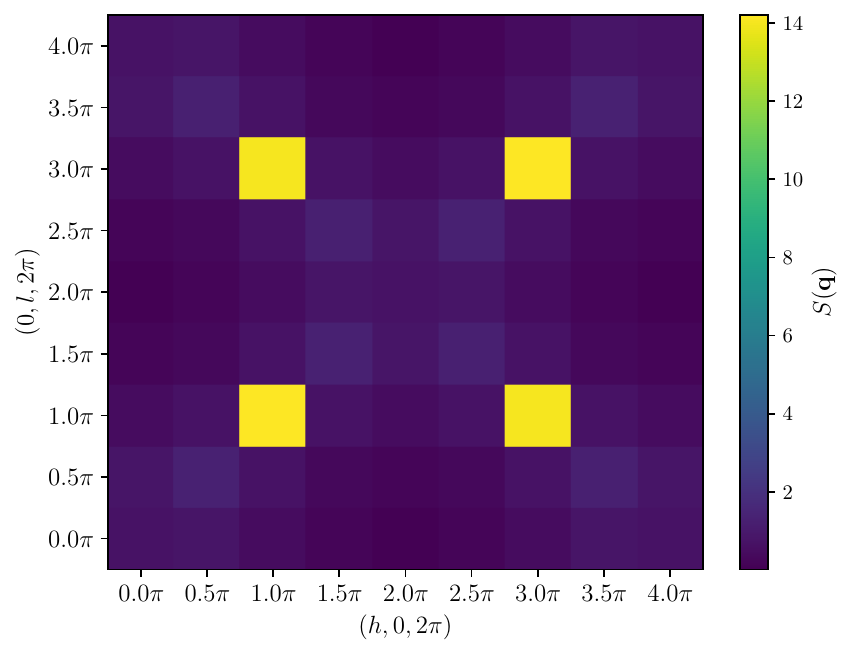} &
  \includegraphics[width=0.3\textwidth]{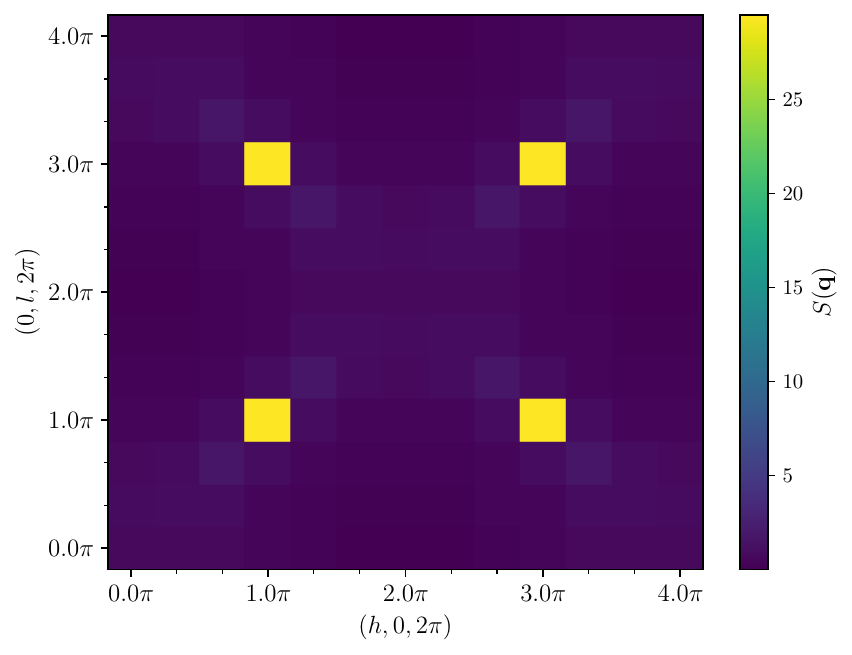} \\
\end{tabular}
\caption{Ground-state structure factors of the tetragonal model for various $J_{2ab}$ and system sizes in the $(h,l,2\pi)$ plane.}
\label{fig:tetragonal_structure_factor}
\end{figure*}
\begin{figure*}
\includegraphics[width=12cm]{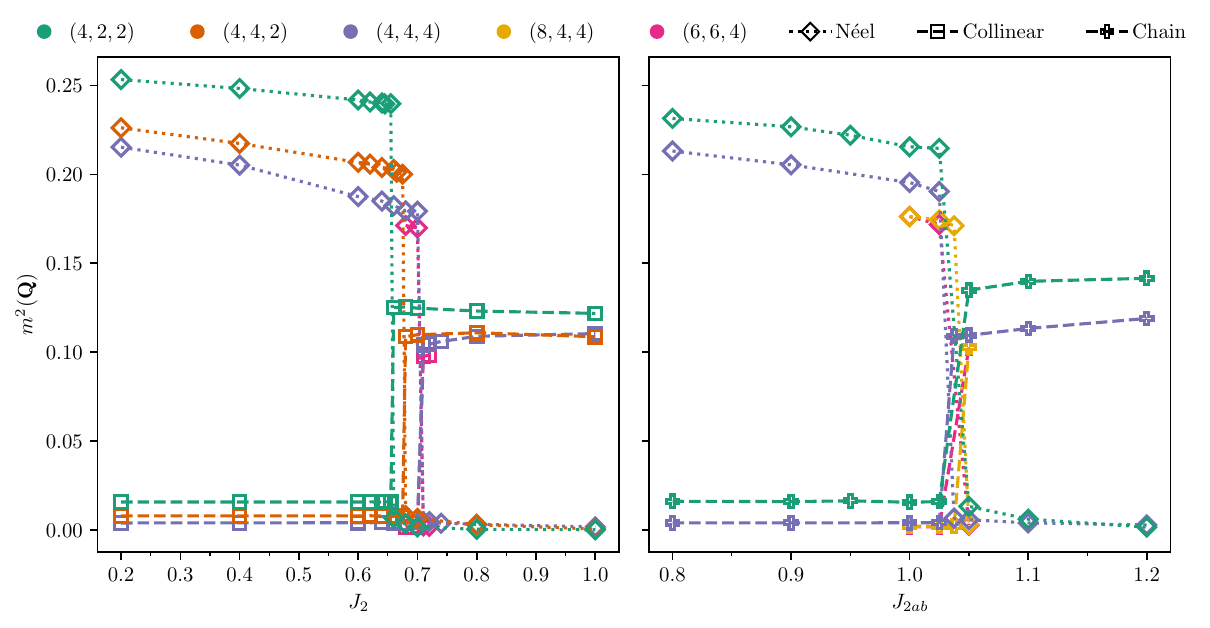}
\caption{Ground-state order parameters for the cubic (left) and tetragonal (right) models for different lattice sizes.}
\label{fig:order_parameters}
\end{figure*}

Static structure factors (Eq.~\ref{eq:structure_factor}) for the ground states we obtain for the cubic and tetragonal models are shown in Fig.~\ref{fig:cubic_structure_factor} and Fig.~\ref{fig:tetragonal_structure_factor} respectively.
As expected from the correlation ratios, the Bragg peaks of the respective ordered phases sharpen as system size is increased.
We also plot the order parameters,
\begin{equation}
    m^2(\mathbf{Q}) = \frac{S(\mathbf{Q})}{N}
\end{equation}
in Fig.~\ref{fig:order_parameters}, with values of $\mathbf{Q}$ defined in Eq.~\ref{eq:Q}.
\section{Tetragonally-distorted $J_1-J_2-J_3$ Ising model}
Energies for the Néel, collinear and chain phases according to Eq.~\ref{eq:ham_j3} are
\begin{eqnarray}
	\frac{E_{\mathrm{N\acute eel}}}{N} = -4J_1 + J_{2c} + 2J_{2ab} + 4J_{3c} + 2J_{3ab}\\
	\frac{E_{\mathrm{collinear}}}{N} = - J_{2c} -2J_{2ab} + 4J_{3c} + 2J_{3ab}\\
	\frac{E_{\mathrm{chain}}}{N} = J_{2c} - 2J_{2ab} -4J_{3c} + 2J_{3ab},
\end{eqnarray}
such that the Néel and collinear phases are stabilized by a ferromagnetic $J_{3c}$ and the chain phase destabilized.
$J_{3ab}$ equally shifts the energy of all phases.
Thus the phases have the same energy along $J_{3c} = J_{2c}/4= 1/2(J_1 - J_{2ab})$.
So for ferromagnetic $J_{2c}$ and antiferromagnetic $J_{2ab} < J_1$, a ferromagnetic $J_{3c}$ causes these phases to compete.
\clearpage
\bibliography{bibliography}
\end{document}